\documentclass[twocolumn,aps,showpacs,groupedaddress,floatfix,prl]{revtex4-1}
\usepackage{amsmath}
\usepackage{graphicx}
\usepackage{dcolumn}
\usepackage{bm}
\usepackage{verbatim}

\begin{document}
\setlength{\parskip}{0mm}
\title{Friction Scaling Laws for Transport in Bacterial Turbulence}
\author{Sanjay C.P. and Ashwin Joy}
\email{ashwin@iitm.ac.in}
\affiliation{Department of Physics, Indian Institute of Technology Madras, Chennai - 600036, India}
\date{\today}
\begin{abstract}
Understanding the role of frictional drag in diffusive transport is an important problem in the field of active turbulence. Using a continuum model
that applies well to bacterial suspensions, we investigate the role of Ekmann friction on the transport of passive (Lagrangian) tracers that go with
the local flow. We find that the crossover from ballistic to diffusive regime happens at a time scale $\tau_c$ that attains a minimum at zero
friction, meaning that both injection and dissipation of energy delay the relaxation of tracers. We explain this by proposing that $\tau_c \sim 2
\ell^*/u_{\text{rms}}$, where $\ell^*$ is a dominant length scale extracted from energy spectrum peak, and $u_{\text{rms}}$ is a velocity scale that
sets the kinetic energy at steady state, both scales monotonically decrease with friction.  Finally, we predict robust scaling laws for $\ell^*$,
$u_{\text{rms}}$ and the diffusion coefficient $\mathcal{D} \sim \ell^* u_{\text{rms}} / 2$, that are valid over a wide range of fluid friction. Our
findings might be relevant to transport phenomena in a generic active fluid.
\end{abstract} 

\maketitle

From a cup of coffee to the scales of the interstellar medium, fluid turbulence is ubiquitous in nature \cite{frisch1995turbulence,
davidson2015turbulence}.  Often stated as the last unsolved problem of classical physics, turbulence is actually an ensemble of non-linear processes
that usually happens in any fluid which is far from equilibrium. These nonlinear processes are responsible for cascading the fluid energy throughout
the inertial range of wave numbers i.e between the scale of injection where turbulence is excited to the scale of damping where viscosity plays a
dominant role \cite{kraichnan1967inertial,boffetta2012two}. The large number of degrees of freedom that are coupled by these nonlinear processes
display highly irregular dynamics and impose a pressing need for a statistical treatment of turbulence. A natural concern is whether the statistical
predictions made in the inertial range are universally applicable, for instance, will the predicted scaling laws hold for general patterns of energy
injection, transfer, and dissipation? Relevant scenarios include mesoscale turbulence in active ``living'' fluids such as bacterial suspensions,
micro-tubule networks or even artificial swimmers \cite{kessler1997collective,sokolov2007concentration,sokolov2012physical}.  It is therefore very
interesting to see whether established results of classical turbulence can be extended to these active fluids. One problem that naturally forms a
basis for a statistical treatment of turbulence is the transport of passive tracers advected by the local Eulerian flow. In this letter, we report a
detailed study of transport in a distribution of such passive tracers in a two-dimensional (2D) active fluid that well describes dense bacterial
suspensions. We record the mean squared displacements of these passive tracers and find that the crossover time $\tau_c$ from ballistic regime to
diffusive regime has a minimum when the fluid friction is zero. The other two cases, namely, positive friction (energy damping) and
negative friction (energy injection) both tend to progressively increase the crossover time $\tau_c$. We explain this minimum in $\tau_c$ by directly
invoking appropriate scale arguments in the homogeneous steady state of the turbulent fluid.
Finally, we set up a universal friction scaling law for the diffusion co-efficient of these passive tracers that is in excellent agreement with our
direct numerical simulations. In the following, we start by describing the numerical setup used in our work. 

\textit{Numerical Simulations:} The literature on continuum theories of active matter with particular emphasis on collective behavior, is quite
extensive \cite{koch2011collective,thampi2013velocity,ramaswamy2010mechanics,baskaran2009statistical,kruse2004asters,saintillan2008instabilities,
simha2002hydrodynamic}. In our work, we use a minimal phenomenological model that was recently developed to study turbulence in a dense suspension of
bacterium \textit{Bacillus subtilis} \cite{wensink2012meso, bratanov2015new, PhysRevLett.110.228102}.  According to this model, at high enough
concentration, this bacterial suspension can be approximated as an incompressible fluid, whose velocity field is governed by the following
equations,
\begin {align}
\label{active_model}
\frac{\partial \bm u}{\partial t}+\lambda_{0}( \bm u \cdot \bm \nabla) \bm u &= - \bm \nabla P -\Gamma_{0} \bm \nabla^{2} \bm u - \Gamma_{2} \bm
\nabla^{4} \bm u - \mu (u) \bm u, \nonumber \\
\bm \nabla \cdot \bm u &= 0,
\end{align}
\begin{figure}[h]
    \centering
      \includegraphics[width=\textwidth,height=0.25\textheight,keepaspectratio]{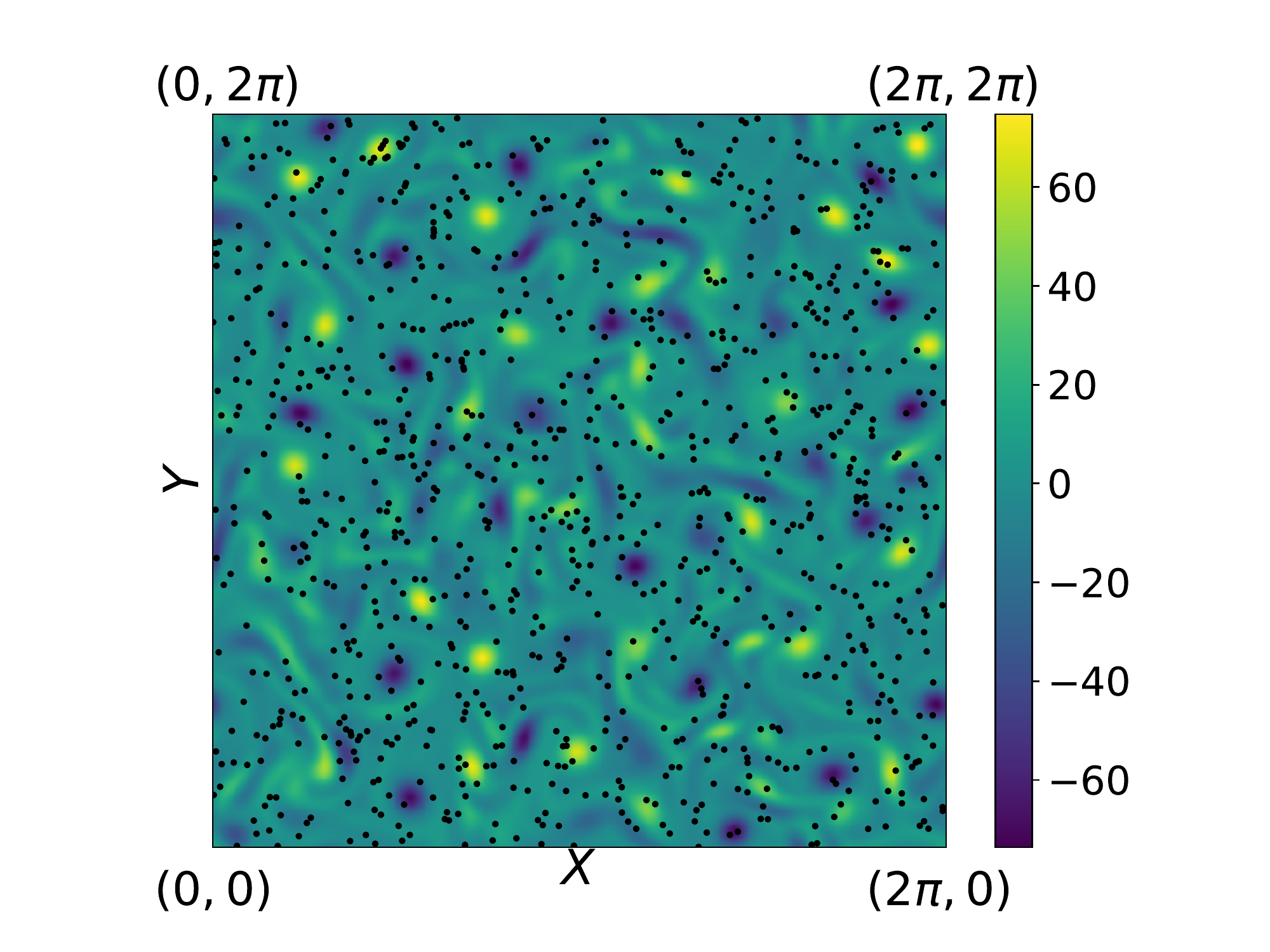}
        \caption{(color online). Color map of the 2D vorticity field along with the distribution of passive tracers at steady state. Grid resolution
          is $512^2$ and Ekmann friction $\alpha = -1$.}
          \label{w-plot}
\end{figure}
where $P$ is pressure and the coefficient of the advective term $\lambda_{0}$ decides the type of bacteria, meaning it is either a \textit{pusher} or
a \textit{puller}, corresponding to the case $\lambda_{0} > 0$ or $\lambda_{0} < 0$, respectively. The model also provides internal drive/ dissipation
through the terms $\Gamma_{0,2}$ and the scalar field $\mu(u) = \alpha + \beta |\bm u|^2$, introduced by Toner and Tu to model the ``flocking''
behavior in self-propelled rod-like objects \cite{swift1977hydrodynamic, toner2005hydrodynamics,toner1998flocks}. Keeping $\Gamma_{0,2} > 0$ leads to
destabilization of a band of wave vectors that mimics energy injection into the bacterial suspension mediated by fluid instabilities.  While the
parameter $\alpha$, hereafter referred to as the Ekmann friction, can produce the effect of energy damping ($\alpha > 0$) as well as energy injection
($\alpha<0$), the parameter  $\beta$ however, is restricted to have only positive values.  In order to non-dimensionalize Eq. (\ref{active_model}), we
normalize all distances to $l_0 = 5\pi/k_0$, where $k_0 = \sqrt{\Gamma_0 / (2\Gamma_2)}$ is the fastest growing mode in the linear regime
characterized by the growth rate: $\gamma(k) = -\alpha + \Gamma_{0}k^{2} -\Gamma_{2}k^{4}$ \cite{bratanov2015new}. Further defining a characteristic
velocity unit, $u_0 = \sqrt{\Gamma_{0}^{3}/\Gamma_{2}}$, one obtains the normalization for time as $t_0 = l_0/u_0$.  In these units, we set the values
of $\Gamma_{0} = (5 \pi \sqrt{2})^{-1}$, $\Gamma_{2} = (5 \pi \sqrt{2})^{-3}$ and $\lambda_0 = 3.5$ to remain consistent with earlier works
\cite{wensink2012meso, bratanov2015new}. All physical quantities henceforth appearing in this letter are expressed in these reduced units.  Equation
(\ref{active_model}) is numerically solved through a pseudo-spectral approach \cite{canuto2012spectral} using a grid with $512^{2}$ points and a
periodic square box of size $[0, 2\pi] \times [0, 2\pi]$.  Starting from random vortices, we time evolve the Eq. \ref{active_model} using the
Crank-Nicholson scheme with a time step $\Delta t = 2 \times 10^{-4}$. Numerical stability is guaranteed by satisfying the Courant-Friedrichs-Lewy
criterion, $(u_x / \Delta_x + u_y / \Delta_y) \Delta t \leq 1$, where $u_{x,y}$ and $\Delta_{x,y}$ are respectively, the components of the velocity
field and grid spacings in two dimensions. Transport properties are further studied by adding $N = 10,000$ tracer particles that just go with the
local flow, and with dynamics,
\begin{equation}
\frac{\text{d}\bm x}{\text{d}t} = \bm u(\bm x,t),
\end{equation}
\begin{figure}[h]
      \centering
      \includegraphics[width=\textwidth,height=0.25\textheight,keepaspectratio]{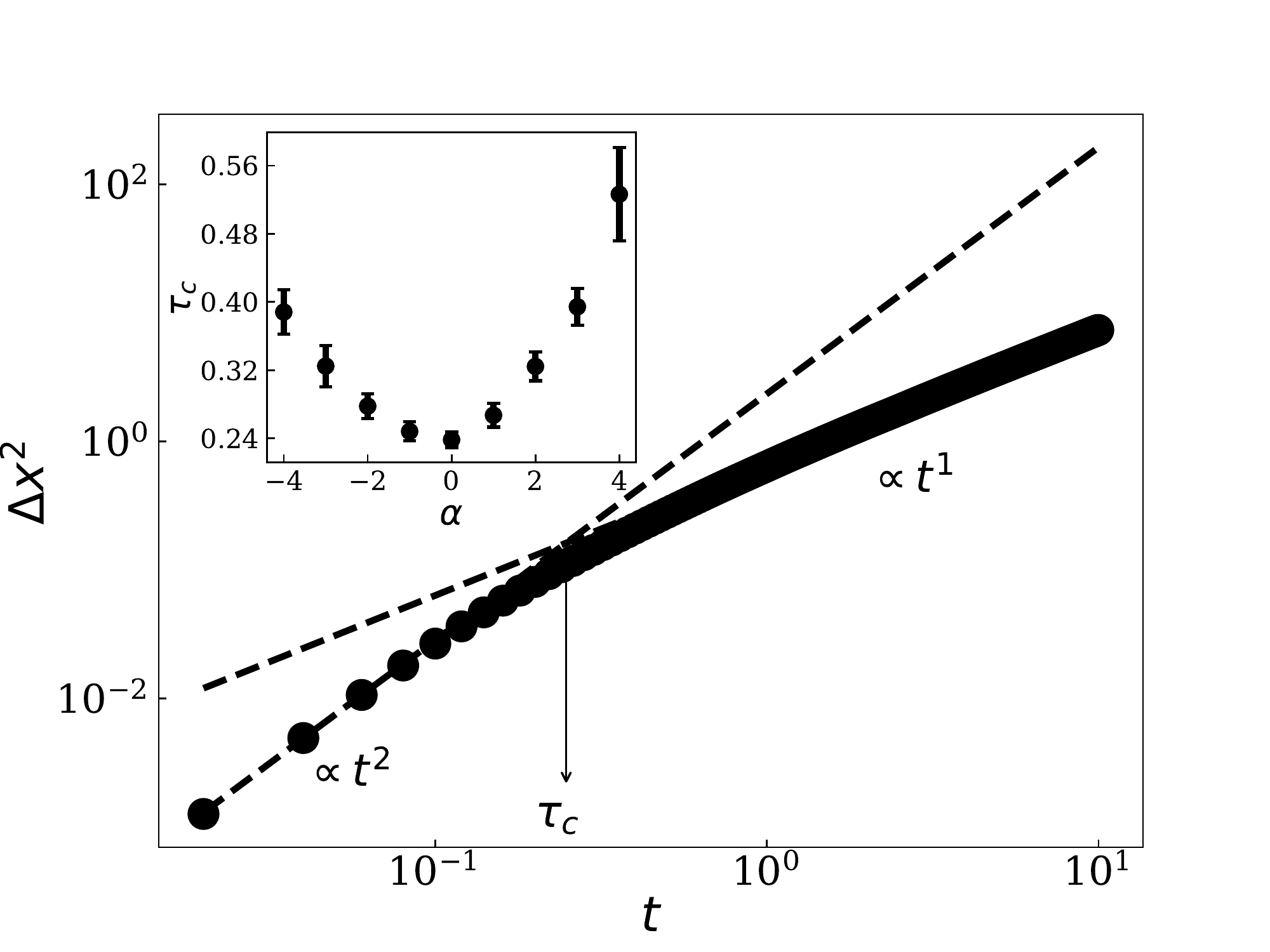}
      \caption{(color online). Mean square displacement $\Delta \bm x^2$ vs. time $t$ for a typical Ekmann friction $\alpha$. Inset: Cross over time $\tau_{c}$ 
       develops a minimum at zero friction (see text for details). Error bars indicate one standard deviation.}
          \label{tau_c-alpha}
\end{figure}
\begin{figure}[h]
     \centering
     \includegraphics[width=\textwidth,height=0.25\textheight,keepaspectratio]{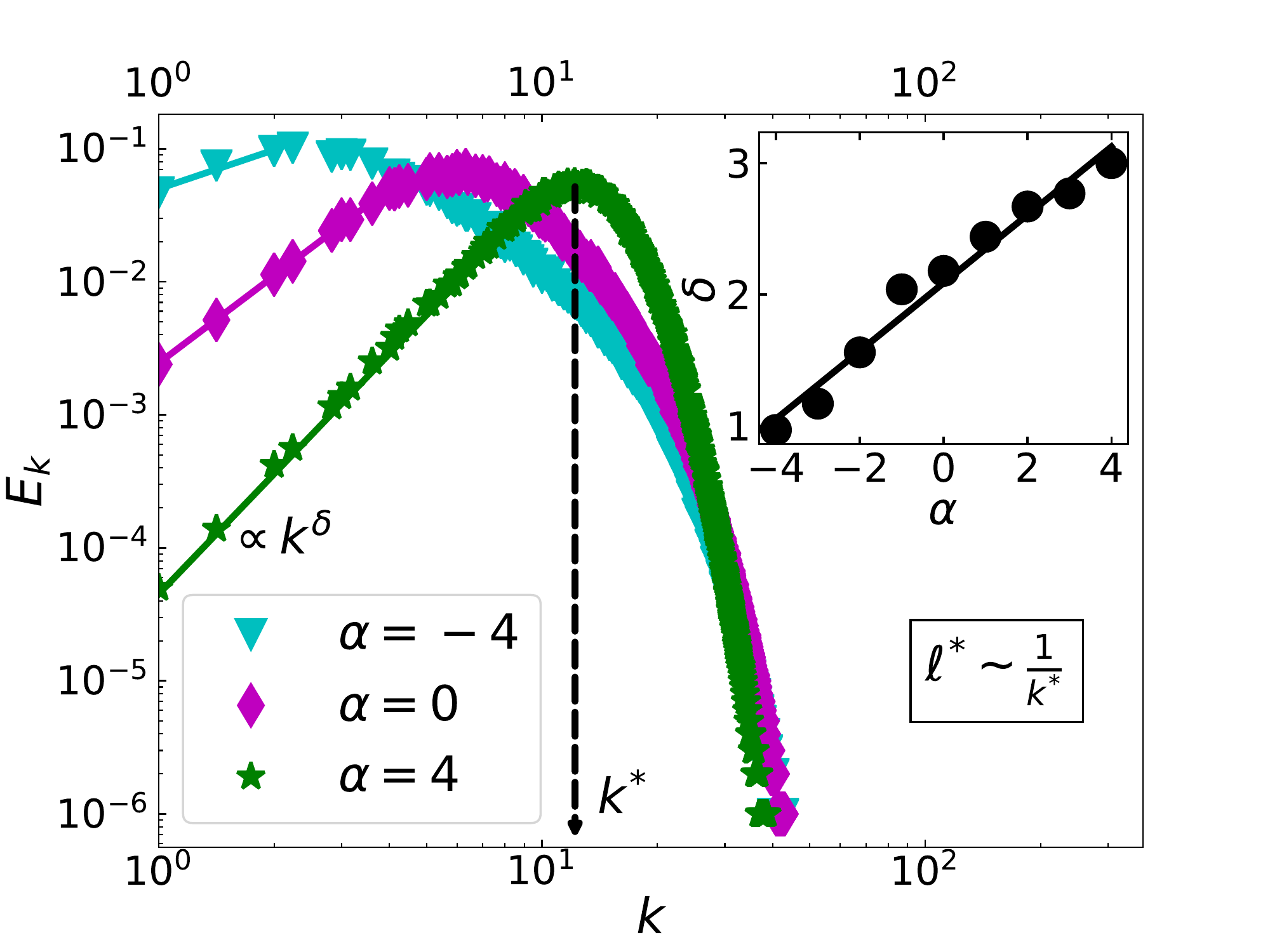}
     \caption{(color online). Energy spectrum $E_k$ for different values of Ekmann friction $\alpha$. The peak shows the location of dominant length scale $k^* =
     1/\ell^*$. Inset: The exponent $\delta$ at low wave numbers is seen to be linear in $\alpha$.} 
     \label{length-scale}
\end{figure}
where $\bm x$ is the position of the tracer particle and $\bm u(\bm x,t)$ gives velocity field of the fluid at $\bm x$, which is found using a cubic
spline interpolation. Figure \ref{w-plot} shows the snapshot of tracer particles distributed in the steady state  turbulent vorticity field. In what
follows, we present a systematic report of our observations on tracer transport and how it scales with Ekmann friction.  We first compute the mean
square displacement of the tracer particles defined as, \begin{equation} \Delta \bm x^2  = \biggl \langle\frac{1}{N} \sum_{i=1}^N \;|\bm x_i(t) -\bm
  x_i(0) |^{2} \biggr \rangle, \end{equation} where $\langle \cdot \rangle$ denotes an ensemble average over 75 independent initial conditions. As is
evident from Fig. \ref{tau_c-alpha}, there is a ballistic regime at $t \ll \tau_c$, where due to correlations in velocity field, we observe $\Delta
\bm x^2\sim t^{2}$. At late times, i.e $t \gg \tau_c$, we observe a diffusive regime where $\Delta \bm x^2\sim t$, due to loss of memory.  The cross
over from ballistic to diffusive regime happens at an intermediate time scale, $\tau_c$ that essentially denotes the relaxation time of the passive
tracers. 
\begin{figure*}
  \begin{minipage}[b]{.48\textwidth}
   \includegraphics[width=\textwidth,height=0.25\textheight,keepaspectratio]{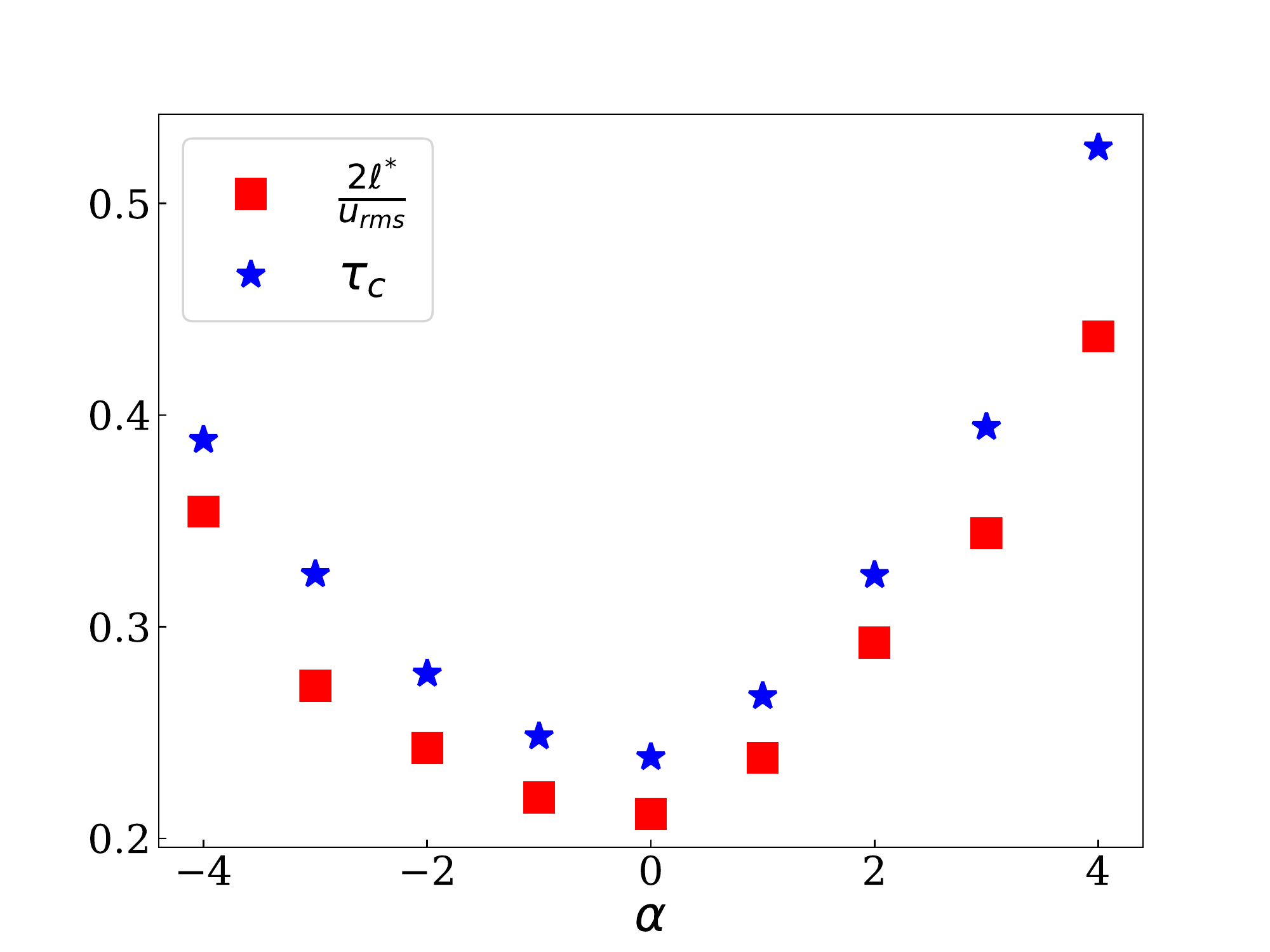}
   \caption{(color online). Comparison of time scales $\tau_c$ and $2 \ell^{*}/u_{\text{rms}}$ extracted respectively from $\Delta \bm x^2$ and
   $E_k$ data.}
   \label{l_by_u_fig}
 \end{minipage}\qquad
 \begin{minipage}[b]{.48\textwidth}
   \includegraphics[width=\textwidth,height=0.25\textheight,keepaspectratio]{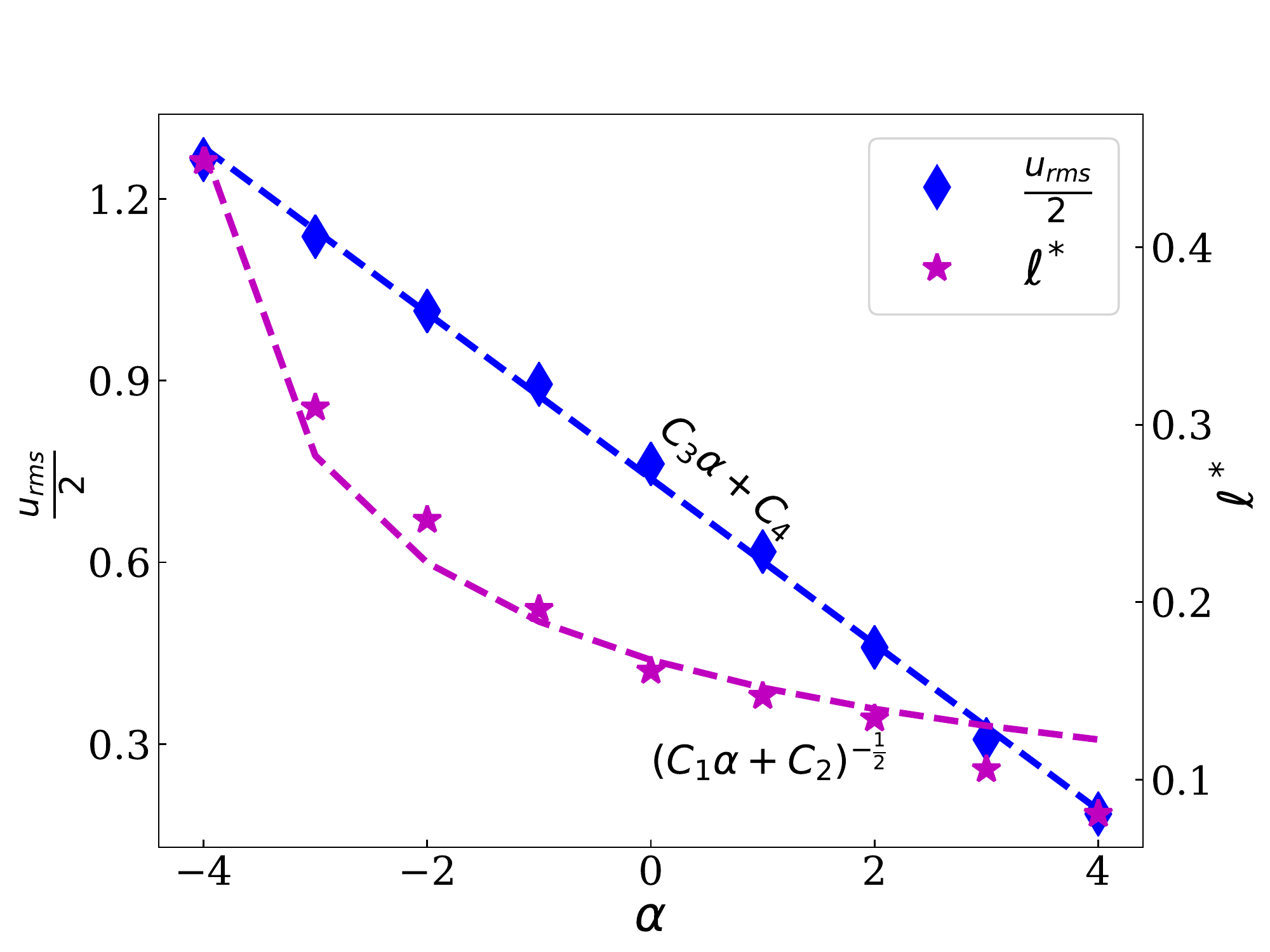}
   \caption{(color online). Characteristic scales $\ell^*$ and $u_{\text{rms}}$ along with their
 predicted scalings shown as dashed lines.}
 \label{l_and_u_fig}
 \end{minipage}
 \end{figure*}
Interestingly, we observe that $\tau_{c}$ develops a minimum at $\alpha = 0$, directly implying that the fastest relaxation occurs at zero
friction (see Fig. \ref{tau_c-alpha} inset). To understand this non-monotonic behavior, we propose to identify $\tau_c \sim 2 \ell^{*}/u_{\text{rms}}$
where the length scale $\ell^{*} = 1/k^*$ with $k^*$ being the location of the peak of the energy spectrum (refer Fig. \ref{length-scale}) and the
velocity scale $u_{\text{rms}} = \sqrt{\langle |\bm u|^2\rangle}$ in the steady state. In Fig. \ref{l_by_u_fig} we show that both these time scales are
in close agreement and display a minimum at $\alpha = 0$, clearly justifying our choice. A physical reasoning for this observed
minimum can be constructed by noticing that $1/|\alpha|$ is a time scale over which advective terms necessary to develop turbulence, are damped.  If we
further realize that $l_0/u_0$ is a time scale over which the unstable modes grow due to stress induced instabilities, the necessary condition for
developed turbulence can be constructed through an effective Reynolds number
\begin{equation}
  \mathcal{R}_E \sim \frac{u_0}{l_0 |\alpha|} \sim \frac{\Gamma_0^2}{\Gamma_2 |\alpha|} \gg 1.
  \label{Reynolds}
\end{equation}
We see that for $\alpha \rightarrow 0$, the above inequality is automatically established for arbitrary values of the parameters $\Gamma_{0,2}$. Since the
transport of tracers is aided by background turbulence, the crossover from ballistic to diffusive regime must be fastest at zero friction.  In what follows, we provide
scaling laws for both $\ell^{*}$ and $u_{\text{rms}}$ that is further used to set up a friction scaling law for diffusion coefficient $\mathcal{D}$ in
our active fluid.  

Recently, an expression for the energy spectrum of our model fluid was derived in Ref.
\cite{bratanov2015new} as 
\begin{equation}
E_k = E_0 k^\delta \; \text{exp} (-E_1 k^2),  
\label{Ekequation}
\end{equation}
where the constants $E_{0,1}$ depend on model parameters excluding $\alpha$. The location
of the peak $k^* \sim 1/\ell^*$ can be easily obtained by extremizing
Eq. \ref{Ekequation}, yielding
\begin{equation}
  k^* \sim \sqrt{\delta},
  \label{k*}
\end{equation}
which together with the observation that $\delta$ is linear in $\alpha$ (see Fig.
\ref{length-scale}) leads us to the following scaling
\begin{equation}
  \ell^* \sim (C_1 \alpha + C_2)^{-1/2},
  \label{l-scaling-eq}
\end{equation}
where the constants $C_{1,2}$ depend on other parameters of the governing model. Next we derive the
scaling for $u_{\text{rms}}$ by focusing on kinetic energy of the fluid in the steady state. 
To that end, we rewrite Eq. \ref{active_model} as 
\begin{widetext}
  \begin{equation}
    \frac{\partial \bm u}{\partial t} + \lambda_0 \biggl[ \frac{1}{2} \bm \nabla (\bm u \cdot \bm u) - \bm u \times \bm \omega \biggr] = - \bm \nabla P 
    + \Gamma_0 \bm \nabla \times \bm \omega - \Gamma_2 \bm \nabla \times \bm \nabla \times \bm \nabla \times \bm \omega - \alpha \bm u - \beta
    \lvert \bm u \rvert^2 \bm u,
    \label{active_model_2}
  \end{equation}
and by taking a scalar product of $\bm u$ with Eq. \ref{active_model_2} we arrive at the energy equation 
    \begin{equation}
      \frac{\partial}{\partial t} \frac{\lvert \bm u \rvert^2}{2} + \frac{\lambda_0}{2} \bm \nabla \cdot (|\bm u|^2 \bm u) = -\bm \nabla \cdot
      (P \bm u) - \Gamma_0 [\bm \nabla \cdot (\bm u \times \bm \omega) - \lvert \bm \omega \rvert^2] - \Gamma_2 [\bm \nabla \cdot ((\bm \nabla \times \bm \nabla
      \times \bm \omega) \times \bm u) - (\bm \nabla^2 \bm \omega) \cdot \bm \omega] - \alpha \lvert \bm u \rvert^2 - \beta \lvert \bm u\rvert^4.
      \label{energy_eq}
        \end{equation}
\end{widetext}
\noindent
The approximation of statistical homogeneity and spatial isotropy allows us to perform a spatial
(ensemble) average of the above equation and put all divergences to zero, leading us to
\begin{equation}
  \frac{\text{d}}{\text{d}t} \frac{\langle \lvert \bm u \rvert^2 \rangle}{2} = \Gamma_0 \langle \bm \omega^2 \rangle + \Gamma_2 \langle (\bm \nabla^2 \bm \omega) \cdot \bm \omega
  \rangle - \alpha \langle \lvert \bm u \rvert^2 \rangle  -  \beta \langle \lvert \bm u\rvert^4 \rangle 
  \label{energy_eq_steady}
\end{equation}
Further under conditions of steady state we drop the time derivative, and by using the scalings 
\begin{align}
  \langle \omega^2 \rangle \sim \frac{u_{\text{rms}}^2} {{\ell^*}^2} \nonumber \\
  \langle (\bm \nabla^2 \bm \omega) \cdot \bm \omega \rangle \sim \frac{u_{\text{rms}}^2}{{\ell^*}^4}
  \label{scalings_2}
\end{align}
in equation \ref{energy_eq_steady}, we are led to 
\begin{equation}
  0 = \frac{\Gamma_0 u_{\text{rms}}^2}{{\ell^*}^2} + \frac{\Gamma_2 u_{\text{rms}}^2}{{\ell^*}^4} - \alpha u_{\text{rms}}^2 -\beta u_{\text{rms}}^4
  \label{scalings_3}
\end{equation}
which together with the scaling for $\ell^*$ (refer Eq. \ref{l-scaling-eq}), yields
\begin{equation}
  \frac{u_{\text{rms}}}{2} \sim C_3 \alpha + C_4
  \label{u-scaling-eq}
\end{equation}
where the constants $C_{3,4}$ again depend on other parameters of the governing model. In Fig.
\ref{l_and_u_fig}, we present a direct comparison of
these scalings (Eq. \ref{l-scaling-eq}, \ref{u-scaling-eq}) with the numerically extracted values of
$\ell^* = 1/k^*$ and $u_{\text{rms}} = \sqrt{\langle |\bm u|^2 \rangle}$ and observe an excellent
agreement between them. 
\begin{figure}[]
  \centering
\includegraphics[width=\textwidth,height=0.25\textheight,keepaspectratio]{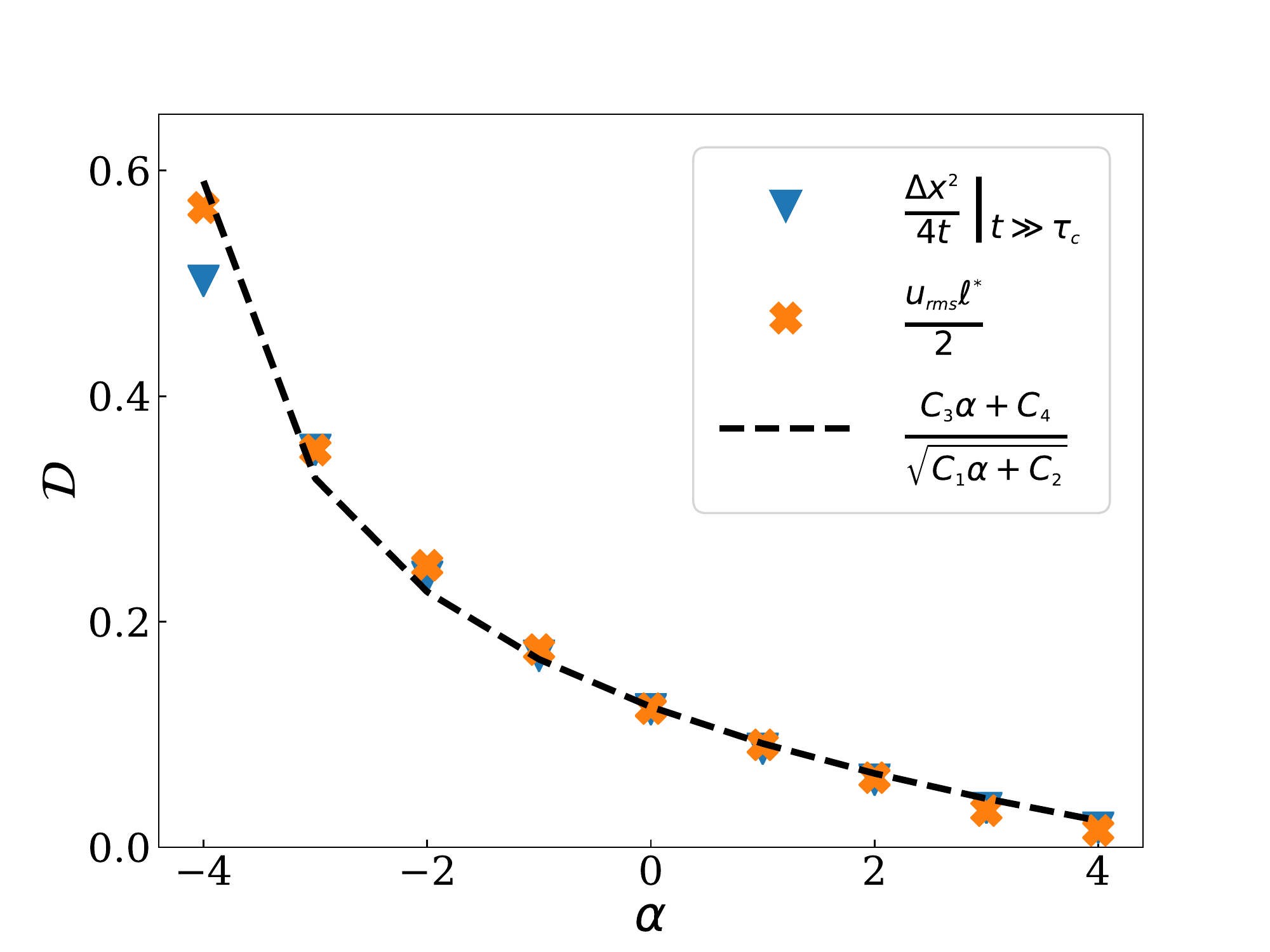}
  \caption{Diffusion coefficient $\mathcal{D}$ vs. Ekmann friction $\alpha$. The clear agreement between the estimates, namely, the asymptotic value
  of $\Delta \bm x^2 / 4 t$ and $u_{\text{rms}} \ell^* /2$ asserts the validity of our scaling law shown here as a dashed line.}
  \label{D-figure}
\end{figure}
To assert the quality of our prediction, we will now present a scaling law for the
tracer diffusion coefficient that remains valid throughout the range of Ekmann friction explored in our
work. We start with the time derivative of the mean squared displacement
\begin{equation}
  \frac{\text{d}}{\text{d} t} \Delta \bm x^2 = 2 \Delta \bm x  \cdot \bm u(t) = 2 \bm u(t) \cdot \int_0^t \bm u(t') \;\text{d}t'
  \label{msd-dot}
\end{equation}
which under the change of variable $\tau = t - t'$ yields
\begin{equation}
  \frac{\text{d}}{\text{d} t} \Delta \bm x^2 = 2 \int_0^t \bm u(t) \cdot \bm u (t-\tau) \;\text{d}\tau. 
  \label{msd-dot-2}
\end{equation}
and in the diffusive regime ($t \gg \tau_c$), further reduces to
\begin{equation}
  \frac{\text{d}}{\text{d} t} \Delta \bm x^2 \approx 2
  u_{\text{rms}} \ell^* \approx 4 \mathcal{D}
  \label{msd-dot-2}
\end{equation}
where the last approximation follows from Einstein's relation. Combining the above equation and scalings for  $\ell^*$ and $u_{\text{rms}}$ derived
respectively in Eqs. \ref{l-scaling-eq} and \ref{u-scaling-eq}, we present below a scaling formula for diffusion coefficient  
\begin{equation}
  \mathcal{D} \approx u_{\text{rms}} \ell^* /2 \sim \frac{C_3 \alpha + C_4}{\sqrt{C_1 \alpha + C_2}}
  \label{D-scaling-eq}
\end{equation}
In Fig. \ref{D-figure}, we show the estimates for diffusion coefficient obtained from two different methods and also plot the scaling law for
comparison. We clearly see an excellent agreement between the $\mathcal{D}$ estimated from the asymptotic value of $\Delta \bm x^2 / 4 t$ and the scale combination
$u_{\text{rms}} \ell^* /2$. It is therefore not surprising that the scaling law presented in Eq. \ref{D-scaling-eq} agrees well with the
foregoing estimates directly obtained from simulation and remains valid upto extreme values of friction (i.e. $\alpha = \pm 4$) permissible by
the conditions of numerical stability of our simulation.

We have investigated in detail the problem of turbulent transport of  passive (Lagrangian) tracers in a model active fluid that well describes dense
bacterial suspensions. We show that in the fully developed turbulence, the tracer transport exhibits a crossover from ballistic to diffusive regime at
a time scale $\tau_c$ that attains a minimum at zero friction. We explain this observation by measuring a characteristic length scale $\ell^*$
extracted from the energy spectrum peak and a characteristic velocity scale $u_{\text{rms}}$ extracted from the total kinetic energy in the steady
state.  Further, by assuming statistical homogeneity and isotropy in the steady state we are able to provide scaling laws for both $\ell^*$ and
$u_{\text{rms}}$ that hold well over a  wide range of Ekmann friction, namely, from the regime of strong energy damping to the regime of strong energy
injection. Finally, we use these characteristic scales to develop a friction scaling law for the diffusion coefficient $\mathcal{D}$. The reported
laws are universal and are not affected by changes in the parameters of the governing model (not shown here) and should apply well to any generic
fluid not limited to bacterial suspensions alone. The findings of our work should also draw the attention of fluid dynamics researchers who are
interested in pursuing the statistical laws that remain universal over general patterns of energy injection and dissipation.

\begin{acknowledgments}
We thank Abhijit Sen and Gautam Menon for discussions and comments on the manuscript. All simulations were done on the VIRGO super cluster of IIT
Madras.
\end{acknowledgments}

\bibliography{bacterial-turbulence}
\end{document}